\documentclass[twocolumn,secnumarabic,amssymb, nobibnotes, aps, pra]{revtex4-1}

\usepackage{amsmath}
\usepackage{amsfonts}
\usepackage{amssymb}
\usepackage{graphicx}
\usepackage{braket}
\usepackage{tikz}
\usepackage[left=1.2cm,right=1cm,top=2cm,bottom=2cm]{geometry}

\def\deltaw{\mathop{\mbox{\normalfont $\delta w$}}\nolimits}

\def\lambdaqft{\mathop{\mbox{\normalfont $\lambda_{\tiny{\text{FT}}}$}}\nolimits}
\usepackage{graphicx}
\usepackage{dcolumn}
\usepackage{bm}

\begin{document}

\preprint{AIP/123-QED}

\title{A frequency-coded QKD scheme with an extension to qu-quarts}

\author{J. Bonetti}
\author{S. M. Hernandez}
\affiliation{Instituto Balseiro, Av. Bustillo km 9.500, Bariloche (R8402AGP), Argentina}

\author{D. F. Grosz}
\affiliation{Instituto Balseiro, Av. Bustillo km 9.500, Bariloche (R8402AGP), Argentina}
\affiliation{Consejo Nacional de Investigaciones Cient\'ificas y T\'ecnicas (CONICET), Argentina}

\begin{abstract}
We propose a novel scheme to implement the BB84 quantum key distribution (QKD) protocol in optical fibers based on a quantum frequency-translation (QFT) process. Unlike conventional QKD systems, which rely on photon polarization/phase to encode qubits, our proposal utilizes photons of different frequencies. Qubits are thus expected to reach longer propagation distances due to the photon frequency state being more robust against mechanical and/or thermal fluctuations of the transmitting medium. Finally, we put forth an extension to a security-enhanced four-character-alphabet (qu-quarts) QKD scheme. 
 
\end{abstract}
\maketitle

\section{Introduction}
Quantum cryptography is amongst the most promising applications of quantum mechanics~\cite{gisin2002quantum}, since it enables privacy and inherently secure communication. Due to its simplicity, the BB84 protocol ~\cite{bennett1984quantum,fox2006quantum} is the most well-known single-particle QKD scheme and many BB84 experiments have been successfully carried out along optical fiber channels {\cite{korzh2015provably,hiskett2006long,namekata2005quantum}. However, unwanted effects that corrupt qubits \cite{gisin2002quantum} hinder reaching long propagation distances. For instance, birefringence may change photon polarization \cite{gisin1995statistics} and mechanical and/or thermal fluctuations may modify its phase \cite{bouwmeester2000physics}, seriously limiting quantum-communication efficiency. A very clever solution to tackle these problems relies on encoding qubits by means of their frequency \cite{bloch2007frequency,gabdulhakov2017frequency,huntington2004components}. That is, instead of using two orthogonal polarization or phase states, utilizing two different frequencies. This scheme involves at least two advantages: on the one hand, a photon traveling through an optical fiber is more likely to experience polarization and/or phase changes than changes in its frequency (in general, the latter are caused by design), rendering this type of encoding more stable; on the other hand, the possibility of augmenting the dimension of the encoding space presents itself naturally, as more than two frequencies can be used to implement QKD schemes based on qudits~\cite{rohit2016high}. 


In the standard BB84 protocol~\cite{bennett1984quantum} keys are distributed using qubits prepared in one of two bases, such that an absolutely certain quantum state measured in one of them exhibits maximal uncertainty when measured in the other (e.g., two orthogonal polarization states versus the same base rotated by $\pi/4$). To begin with, we will illustrate our proposal by applying it to the standard BB84 protocol, i.e., by frequency coding qubits, and then we will move to the more involved case of quantum quarts~\cite{bechmann2000quantum} (qu-quarts) based QKD.
We propose to generate the aforementioned frequency-coded qubits relying on a special Four-Wave Mixing (FWM) process known as Quantum Frequency Translation (QFT), or Bragg Scattering (BS)~\cite{kumar1990quantum,mckinstrie2005translation,mcguinness2010quantum}. It consists of a dual-pump configuration FWM (at frequencies $w_{p1}$ and $w_{p2}$), with a small signal at $w_{s1}$ (see Fig.~\ref{fig1}). Interaction between these three frequencies in a nonlinear medium with third-order susceptibility $\chi^{(3)}$~\cite{agrawal2007nonlinear} produces an idler signal at $w_{s2}=w_{p2}-w_{p1}+w_{s1}$. From a quantum mechanical perspective, this process consists of two photon annihilations, one at $w_{p2}$ and another at $w_{s1}$, and the simultaneously creation of two photon at $w_{p1}$ and $w_{s2}$. The opposite process (creations at $w_{p2}$ and $w_{s1}$ and annihilations at $w_{p1}$ and $w_{s2}$) is also allowed in this configuration. Although there are many other possible processes (other creation-annihilation combinations that conserve energy and number of photons, e.g. phase conjugation and modulation instability), they can be neglected if we assume, as we will, that this is the only frequency combination that also satisfies the phase-matching condition \cite{mcguinness2010quantum}  $k_{p1}+k_{s2}=k_{p2}+k_{s1}$. Each photon annihilation at $w_{s1}$ involves a photon creation in $w_{s2}$ and vice versa, so the quantity of photons at $w_{s1}$ or $w_{s2}$ remains constant. When the phase-matching condition is perfectly satisfied, all $n$ photons in the small signal at $w_{s1}$ are annihilated, and $n$ photons at the idler $w_{s2}$ are created. The propagation distance at which this occurs is called the \textit{frequency-translation length} ($\lambdaqft$).

\begin{figure}
\centering
\begin{tikzpicture}
\draw[thick,->] (7,0) -- (12,0);
\node at (11.7,-0.2) (jugh) {$w$};
\draw[thick,->] (10,0) -- (10,3);
\node at (10.4,2.7) (fhnei){$w_{p1}$};
\draw[thick,->] (11,0) -- (11,3);
\node at (11.4,2.7) (fhnei){$w_{p2}$};
\draw[thick,->,red] (7.3,0) -- (7.3,1.5);
\node at (7.7,1.2) (fhnei){$w_{s1}$};
\draw[thick,->,blue] (8.3,0) -- (8.3,1.5);
\node at (8.7,1.2) (fhnei){$w_{s2}$};
\draw[dashed,<-] (8.3,1.8) arc (45:135:0.8);
\node at (7.8,2.3) (jugh) {\tiny{QFT}};
\draw[thick,<->,dashed,gray] (10,0.3) -- (11,0.3);
\node at (10.5,0.5) (hfn){\small{$\deltaw$}};
\draw[thick,<->,dashed,gray] (7.3,0.3) -- (8.3,0.3);
\node at (7.8,0.5) (hfn){\small{$\deltaw$}};
\draw[thick] (9,-0.2) -- (9.1,0.2);
\draw[thick] (9.1,-0.2) -- (9.2,0.2);
\end{tikzpicture}
\caption{Quantum Frequency Translation by a Four-Wave-Mixing process.}
\label{fig1}
\end{figure}
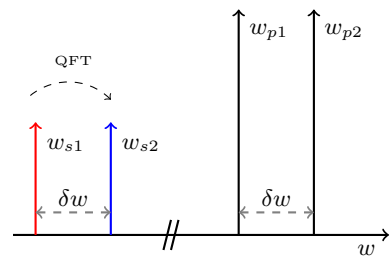
 
If we consider the total QFT, starting with a single photon of frequency $w_{s1}$ at $z=0$ as the initial state, for instance, and obtaining a single photon of frequency $w_{s2}$ at $z=\lambdaqft$ as the final state (see Fig.~\ref{fig2}), a straightforward quantum interpretation of the described FWM process allows us to describe the single-photon QFT process by the photon frequency quantum state  $\ket{\psi}=\mu \ket{w_{s1}} + \nu \ket {w_{s2}}$, where $|\mu|^2$ and $|\nu|^2$ are the  probabilities to find the photon at frequencies $w_{s1}$ and $w_{s2}$, respectively. If the propagation distance is $\lambdaqft$ the photon will have a defined frequency $w_{s2}$ with probability one, i.e., the frequency translation is complete. Between $0$ and $\lambdaqft$, the photon frequency is a quantum superposition of both frequencies $w_{s1}$ and $w_{s2}$.  In fact, as it will be demonstrated in the next section, $|\mu|^2= |\nu|^2 = 0.5$ at $\lambdaqft/2$ thus the photon frequency is maximally uncertain. We call this distance the \textit{half-translation length} ($\lambdaqft/2$).

\begin{figure*}
\centering
\begin{tikzpicture}
\fill[gray!5!white] (0,0)--(2,3.5)--(7,3.5)--(5,0)--(0,0);
\draw[thick,dashed] (0.65,1.2) -- (5.65,1.2);
\draw[thick,dashed] (1.35,2.4) -- (6.35,2.4);
\draw[thick,dashed,gray] (0.8,0)--(2.8,3.5);
\draw[thick,dashed,gray] (2.8,0)--(4.8,3.5);
\node at (1.7,3.4) (jugh) {$z$};
\node at (0.9,2.5) (jugh) {$\lambdaqft$};
\node at (0.2,1.3) (jugh) {$\frac{\lambdaqft}{2}$};
\draw[thick,->] (0,0) -- (5,0);
\node at (4.7,-0.2) (jugh) {$w$};
\draw[thick,->] (0,0) -- (2,3.5);
\draw[thick,->,red] (0.8,0) -- (0.8,1);
\draw[thick,->,red] (1.5,1.2) -- (1.5,1.7);
\draw[thick,->,blue] (3.5,1.2) -- (3.5,1.7);
\draw[thick,->,blue] (4.17,2.4) -- (4.17,3.4);
\node at (0.8,-0.2) (fhnei){$w_{s1}$};
\node at (2.8,-0.2) (fhnei){$w_{s2}$};
\node at (4.65,0.6) (fhnei){$\ket{\psi_1}=\ket{w_{s1}}$};
\node at (6.5,1.7) (fhnei){$\ket{\phi_1}=\frac{1}{\sqrt{2}}\left(\ket{w_{s1}}+i\ket{w_{s2}}\right)$};
\node at (6.1,2.9) (fhnei){$\ket{\psi_2}=\ket{w_{s2}}$};
\fill[gray!5!white] (9,0)--(11,3.5)--(16,3.5)--(14,0)--(9,0);
\draw[thick,dashed] (9.65,1.2) -- (14.65,1.2);
\draw[thick,dashed] (10.35,2.4) -- (15.35,2.4);
\draw[thick,dashed,gray] (9.8,0)--(11.8,3.5);
\draw[thick,dashed,gray] (11.8,0)--(13.8,3.5);
\node at (10.7,3.4) (jugh) {$z$};
\node at (9.9,2.5) (jugh) {$\lambdaqft$};
\node at (9.2,1.3) (jugh) {$\frac{\lambdaqft}{2}$};
\draw[thick,->] (9,0) -- (14,0);
\node at (13.7,-0.2) (jugh) {$w$};
\draw[thick,->] (9,0) -- (11,3.5);
\draw[thick,->,blue] (11.8,0) -- (11.8,1);
\draw[thick,->,red] (10.5,1.2) -- (10.5,1.7);
\draw[thick,->,blue] (12.5,1.2) -- (12.5,1.7);
\draw[thick,->,red] (11.17,2.4) -- (11.17,3.4);
\node at (9.8,-0.2) (fhnei){$w_{s1}$};
\node at (11.8,-0.2) (fhnei){$w_{s2}$};
\node at (13.65,0.6) (fhnei){$\ket{\psi_2}=\ket{w_{s2}}$};
\node at (15.5,1.7) (fhnei){$\ket{\phi_2}=\frac{1}{\sqrt{2}}\left(i\ket{w_{s1}}+\ket{w_{s2}}\right)$};
\node at (15.1,2.9) (fhnei){$\ket{\psi_1}=\ket{w_{s1}}$};
\end{tikzpicture}
\caption{Single-photon QFT (pumps are not showed).}
\label{fig2}
\end{figure*}
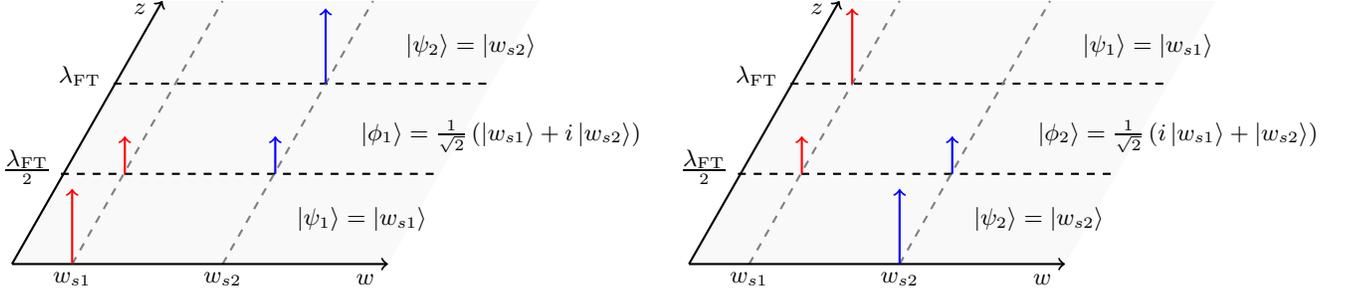

The four states labeled in Fig.~\ref{fig2}  ($\ket{\psi_1}$, $\ket{\psi_2}$, $\ket{\phi_1}$ and $\ket{\phi_2}$) are readily suited to be used as the standard BB84 protocol qubits, where we might name the orthogonal bases $\psi$ and $\phi$. To make and read base $\phi$ qubits, we may propagate a photon across a $\lambdaqft/2$ length QFT system. Alice and Bob decide which base they use, by applying or not this \textit{half-translation} to the photon. When they use different bases, Bob receives a maximally frequency-uncertain photon, like in a standard BB84 scheme.  

\section{Quantitative description of the qubits-based scheme}

A quantum mechanical analysis of the QFT process in optical fibers has been presented in the work of McGuinness et~al.~\cite{mcguinness2011theory}. Treating the pumps as classical fields and the signals as quantum fields, the single-photon frequency state can be described by the equation of motion
\begin{equation}
\label{Schroedinger}
\frac{\partial}{\partial z} \ket{\psi} = i \hat{H} \ket{\psi},
\end{equation}
where the Hamiltonian is~\cite{mckinstrie2005translation}
\begin{equation*}
\hat{H} = \delta \left(\hat{a}^{\dagger}_{s1}\hat{a}_{s1}-\hat{a}^{\dagger}_{s2}\hat{a}_{s2}\right)+\kappa \left(\hat{a}^{\dagger}_{s1}\hat{a}_{s2}+\hat{a}^{\dagger}_{s2}\hat{a}_{s1}\right),
\end{equation*}
 $\hat{a}^{\dagger}$ and $\hat{a}$ are the creation and annihilation operators, respectively, and $\kappa$ is the \textit{effective nonlinearity}, defined as
\begin{equation*}
\kappa = 2\gamma \sqrt{P_1 P_2},
\end{equation*}
where $\gamma$ is the fiber nonlinear coefficient and $P_{1,2}$ are the pump powers. The phase mismatch $\delta$ is given by 
\begin{equation}
\label{phase-matching}
\delta = \frac{\beta_{p2}-\beta_{p1}+\beta_{s1}-\beta_{s2}+\gamma(P_1-P_2)}{2},
\end{equation}
where $\beta_{p1,p2,s1,s2}$ are the propagation constants at each frequency. We choose pump powers such that the process has a perfect phase-matching condition ($\delta =0$).   

By replacing the photon frequency state $\ket{\psi} = \mu \ket{w_{s1}} +\nu \ket{w_{s2}}$ into Eq.~(\ref{Schroedinger}) and solving for $\delta=0$, we obtain
\begin{equation*}
\left( \begin{array}{c} \mu(z) \\ \nu(z) \end{array}\right)=\left(\begin{array}{cc} \cos{(\kappa z)} & i\sin{(\kappa z)}\\  i\sin{(\kappa z)} & \cos{(\kappa z)}\end{array} \right)\left( \begin{array}{c} \mu(0) \\ \nu(0) \end{array}\right).
\end{equation*}                                                                       
This shows that the QFT process has a periodic behavior, switching the probability of finding the photon in one of the two frequencies. Thus, the frequency-translation length can be easily obtained as 
\begin{equation*}
\lambdaqft = \frac{\pi}{2\kappa}.
\end{equation*}

Figure \ref{fig4} shows schematically how to implement the BB84 protocol by the QFT method. Bit values (1 and 0) are coded in the photon frequency. Alice has two single-photon sources (SPS1 and SPS2) of frequencies $w_{s1}$ and $w_{s2}$. By optically switching, she chooses randomly the path that the photon will take. In one of these paths, the photon is sent without changing its frequency state (i.e., Alice uses the base $\psi$). The other path, instead, involves a half-QFT process on the photon and the frequency state changes to a maximally uncertain one (i.e., Alice uses the base $\phi$) when measured in the $\psi$ base. Bob also has an optical switch to choose the measurement base. If they choose the same base, the photon reaches Bob's single-photon detectors (SPD) with a perfectly certain frequency state (i.e., $|\mu|^2 = 1$ and $|\nu|^2 = 0$, or $|\mu|^2 = 0$ and $|\nu|^2 = 1$, depending on whether the bit value was). If they select opposite bases, the photon arrives at the SPD in a maximally uncertain frequency state (i.e., with respect to the measurement base $\psi$) and Bob measures one frequency or the other with the same probability; since this reading provides no information it will eventually be discarded by the QKD protocol.

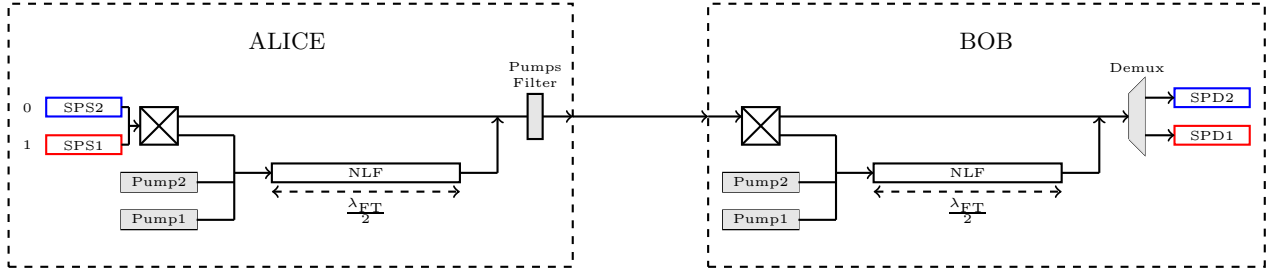
\begin{figure*}
\centering
\begin{tikzpicture}
\draw[thick] (0,0) rectangle (1,0.25);
\draw[thick] (0,0.5) rectangle (1,0.75);
\draw[thick,red] (-1,1) rectangle (0,1.25);
\draw[thick,blue] (-1,1.5) rectangle (0,1.75);
\fill[gray!20!white] (0,0) rectangle (1,0.25);
\fill[gray!20!white] (0,0.5) rectangle (1,0.75);
\node at (0.5,0.12) (jghk){\tiny{Pump1}};
\node at (0.5,0.62) (jghk){\tiny{Pump2}};
\node at (-0.5,1.12) (jghk){\tiny{SPS1}};
\node at (-0.5,1.62) (jghk){\tiny{SPS2}};
\node at (-1.25,1.12) (jghk){\tiny{1}};
\node at (-1.25,1.62) (jghk){\tiny{0}};
\draw[thick] (0.25,1.125) rectangle (0.75,1.625);
\draw[thick] (0.25,1.125)--(0.75,1.625);
\draw[thick] (0.25,1.625)--(0.75,1.125);
\draw[thick] (0,1.625)--(0.1,1.625);
\draw[thick] (0,1.125)--(0.1,1.125);
\draw[thick] (0.1,1.625)--(0.1,1.125);
\draw[thick,->] (0.1,1.375)--(0.25,1.375);
\draw[thick,->] (0.75,1.5)--(6,1.5);
\draw[thick] (0.75,1.25)--(1.5,1.25);
\draw[thick] (1,0.625)--(1.5,0.625);
\draw[thick] (1,0.125)--(1.5,0.125);
\draw[thick] (1.5,1.25)--(1.5,0.125);
\draw[thick,->] (1.5,0.75)--(2,0.75);
\draw[thick] (2,0.625) rectangle (4.5,0.875);
\draw[thick] (4.5,0.75)--(5,0.75);
\node at (3.25,0.75) (jghk){\tiny{NLF}};
\draw[thick,<->,dashed] (2,0.5) -- (4.5,0.5);
\node at (3.25,0.25) (jghk){\tiny{$\frac{\lambdaqft}{2}$}};
\draw[thick,->] (5,0.75)--(5,1.5);
\draw[thick,dashed] (-1.5,-0.5) rectangle (6,3);
\draw[thick,dashed] (7.8,-0.5) rectangle (15.2,3);
\draw[thick,->] (6,1.5)--(7.8,1.5);
\draw[thick,->] (7.8,1.5)--(8.25,1.5);
\node at (2.2,2.5) (jghk){\small{ALICE}};
\node at (11.5,2.5) (jghk){\small{BOB}};
\draw[thick] (8,0) rectangle (9,0.25);
\draw[thick] (8,0.5) rectangle (9,0.75);
\fill[gray!20!white] (8,0) rectangle (9,0.25);
\fill[gray!20!white] (8,0.5) rectangle (9,0.75);
\node at (8.5,0.12) (jghk){\tiny{Pump1}};
\node at (8.5,0.62) (jghk){\tiny{Pump2}};
\draw[thick] (8.25,1.125) rectangle (8.75,1.625);
\draw[thick] (8.25,1.125)--(8.75,1.625);
\draw[thick] (8.25,1.625)--(8.75,1.125);
\draw[thick,->] (8.75,1.5)--(13.4,1.5);

\draw[thick] (8.75,1.25)--(9.5,1.25);
\draw[thick] (9,0.625)--(9.5,0.625);
\draw[thick] (9,0.125)--(9.5,0.125);
\draw[thick] (9.5,1.25)--(9.5,0.125);
\draw[thick,->] (9.5,0.75)--(10,0.75);
\draw[thick] (10,0.625) rectangle (12.5,0.875);
\draw[thick] (12.5,0.75)--(13,0.75);
\node at (11.25,0.75) (jghk){\tiny{NLF}};
\draw[thick,<->,dashed] (10,0.5) -- (12.5,0.5);
\node at (11.25,0.25) (jghk){\tiny{$\frac{\lambdaqft}{2}$}};
\draw[thick,->] (13,0.75)--(13,1.5);

\draw[thick,blue] (14,1.375+0.25) rectangle (15,1.625+0.25);
\node at (14.5,1.75) (jghk){\tiny{SPD2}};

\draw[thick,red] (14,1.375-0.25) rectangle (15,1.625-0.25);
\node at (14.5,1.25) (jghk){\tiny{SPD1}};

\draw[thick,->] (13.6,1.75)--(14,1.75);
\draw[thick,->] (13.6,1.25)--(14,1.25);

\fill[gray!20!white] (5.4,1.2) rectangle (5.6,1.8);
\draw[thick] (5.4,1.2) rectangle (5.6,1.8);
\node at (5.5,2.15) (jhjhm){\tiny{Pumps}};
\node at (5.5,1.95) (jhjhm){\tiny{Filter}};

\draw[thick] (13.4,1.2)--(13.4,1.8)--(13.6,2)--(13.6,1)--(13.4,1.2);
\draw[fill,gray!20!white] (13.4,1.2)--(13.4,1.8)--(13.6,2)--(13.6,1)--(13.4,1.2);
\node at (13.5,2.15) (jhjhm){\tiny{Demux}};

\end{tikzpicture}
\caption{BB84 implementation via QFT. NLF stands for nonlinear fiber.}
\label{fig4}
\end{figure*}

It is important to mention that the proposed QKD scheme is realizable using currently available technologies. 
Recently, frequency translation of single-photons was proven succesful in a photonic crystal fiber~\cite{mcguinness2010quantum} (PCF). Also, single-photon sources are available.

\section{Extension to qu-quarts QKD}

Another significant advantage of frequency encoding is the possibility of having single-photon quantum states represented in a Hilbert space of dimension greater than 2. This fact can be exploited to encode qudits \cite{rohit2016high} (instead of qubits) in each photon. In particular, we put forth an original configuration to produce single-photon states belonging to a four dimensional Hilbert space (i.e., 4 different signal frequencies) by a QFT process similar to that shown in the previous section. In Fig.~\ref{fig5} we show the proposed scheme. There we define the \textit{qu-quart region}, a spectral band which includes the four signal frequencies $w_{s1}$, $w_{s2}$, $w_{s3}$ and $w_{s4}$. Alice has four SPS at these frequencies, which she uses to produce four different keys.

To show how this configuration works, let us consider the following example: If Alice launches a photon of frequency $w$ at the input end of the fiber several QFT processes can occur. The interaction among the photon and pumps 1 and 2 yields two possible results: the photon either "jumps" to $w+\deltaw$ or it jumps to $w-\deltaw$. The photon can also interact in a QFT with pumps 2 and 3. In this case, the photon can jump to $w \pm 3\deltaw$. The interaction with pumps 1 and 3 can also occur, and in this case jumps to $w \pm 4\deltaw$ are possible. However, if properly designed, the qu-quart region limits all the possibles jumps to just a few ones. Ideally, the qu-quart region dispersion will be such that the four signal frequencies are perfectly phase-matched among them but highly phase-mismatched with any frequency external to this region. In practice, this may be achieved by means of a dispersion-flattened fiber or a PCF whose dispersion profile is constant in the qu-quart region and changes abruptly in the boundaries. Jumps outside the qu-quart region are thus forbidden and the only possible QFT processes are the ones shown in Fig.~\ref{fig5}. Note that the reason for including pump 3 is to enable the jump between $w_{s1}$ and $w_{s4}$; furthermore, pump 3 power matches that of pump 1, thus making all frequency jumps to be governed by the same nonlinear coefficient $\kappa$.

\begin{figure*}
\centering
\begin{tikzpicture}
\draw[dashed,<->] (3.95,2) arc (45:135:0.65);
\draw[dashed,<->] (4.95,2) arc (45:135:0.65);
\draw[dashed,<->] (2.95,2) arc (45:135:0.65);
\draw[dashed,<->] (5,2.25) arc (55:125:2.7);

\node at (3.5,2.5) (jugh) {\tiny{QFT}};
\draw[thick,->] (0,0) -- (13.1,0);
\node at (12.7,-0.2) (jugh) {$w$};
\draw[thick,-] (6.5,-0.2)--(6.6,0.2);
\draw[thick,-] (6.6,-0.2)--(6.7,0.2);
\node at (4,1.5) (jugh) {$w_{s3}$};
\node at (5,1.5) (jugh) {$w_{s4}$};
\node at (3,1.5) (jugh) {$w_{s2}$};
\node at (2,1.5) (jugh) {$w_{s1}$};
\draw[thick,->,green!50!black] (4,0) -- (4,1.2);
\draw[thick,->,blue] (3,0) -- (3,1.2);
\draw[thick,->,red] (2,0) -- (2,1.2);
\draw[thick,->,yellow!50!black] (5,0) -- (5,1.2);

\draw[line width=0.5mm,->] (9,0) -- (9,1.5+1.5);
\draw[line width=0.5mm,->] (8,0) -- (8,1.35+1.5);
\draw[line width=0.5mm,->] (12,0) -- (12,2.4);
\node at (8,1.55+1.5) (jugh) {$w_{p1}$};
\node at (9,1.7+1.5) (jugh) {$w_{p2}$};
\node at (12,1.6+1) (jugh) {$w_{p3}$};
\draw[thick,<->,dashed] (8,0.3) -- (9,0.3);
\node at (8.5,0.5) (hfn){\small{$\deltaw$}};

\draw[thick,<->,dashed] (9,0.3) -- (12,0.3);
\node at (10.5,0.5) (hfn){\small{$3\deltaw$}};

\draw[thick,<->,dashed] (3,0.3) -- (4,0.3);
\node at (3.5,0.5) (hfn){\small{$\deltaw$}};

\draw[thick,<->,dashed] (1.8,-0.3) -- (5.2,-0.3);
\node at (3.5,-0.6) (hfnf){qu-quart region};
\end{tikzpicture}
\caption{Frequencies configuration for qu-quarts BB84.}
\label{fig5}
\end{figure*}
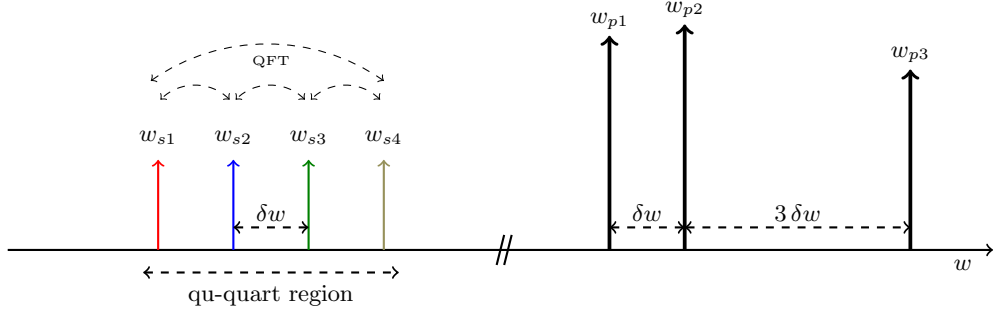

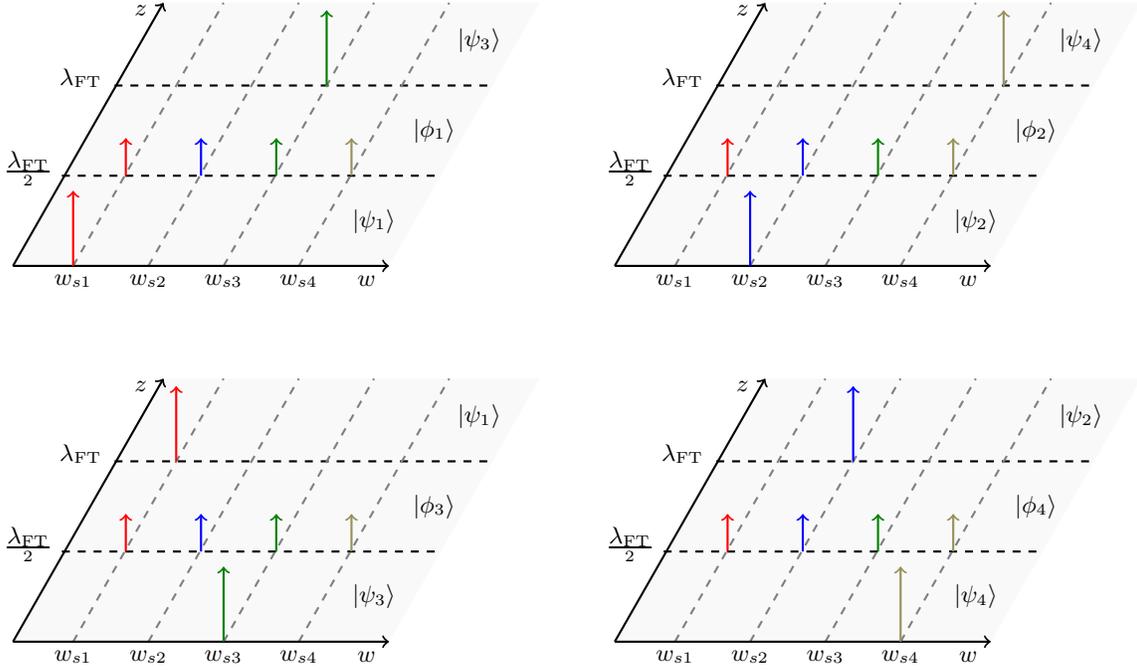
\begin{figure*}
\centering
\begin{tikzpicture}
\fill[gray!5!white] (0,0)--(2,3.5)--(7,3.5)--(5,0)--(0,0);
\draw[thick,dashed] (0.65,1.2) -- (5.65,1.2);
\draw[thick,dashed] (1.35,2.4) -- (6.35,2.4);
\draw[thick,dashed,gray] (0.8,0)--(2.8,3.5);
\draw[thick,dashed,gray] (1.8,0)--(3.8,3.5);
\draw[thick,dashed,gray] (2.8,0)--(4.8,3.5);
\draw[thick,dashed,gray] (3.8,0)--(5.8,3.5);
\node at (1.7,3.4) (jugh) {$z$};
\node at (0.9,2.5) (jugh) {$\lambdaqft$};
\node at (0.2,1.3) (jugh) {$\frac{\lambdaqft}{2}$};
\draw[thick,->] (0,0) -- (5,0);
\node at (4.7,-0.2) (jugh) {$w$};
\draw[thick,->] (0,0) -- (2,3.5);
\draw[thick,->,red] (0.8,0) -- (0.8,1);
\draw[thick,->,red] (1.5,1.2) -- (1.5,1.7);
\draw[thick,->,green!50!black] (3.5,1.2) -- (3.5,1.7);
\draw[thick,->,yellow!50!black] (4.5,1.2) -- (4.5,1.7);
\draw[thick,->,blue] (2.5,1.2) -- (2.5,1.7);
\draw[thick,->,green!50!black] (4.17,2.4) -- (4.17,3.4);
\node at (0.8,-0.2) (fhnei){$w_{s1}$};
\node at (1.8,-0.2) (fhnei){$w_{s2}$};
\node at (2.8,-0.2) (fhnei){$w_{s3}$};
\node at (3.8,-0.2) (fhnei){$w_{s4}$};
\node at (4.8,0.6) (fhnei){$\ket{\psi_1}$};
\node at (5.6,1.8) (fhnei){$\ket{\phi_1}$};
\node at (6.2,3) (fhnei){$\ket{\psi_3}$};

\tikzset{shift={(8,0)}}
\fill[gray!5!white] (0,0)--(2,3.5)--(7,3.5)--(5,0)--(0,0);
\draw[thick,dashed] (0.65,1.2) -- (5.65,1.2);
\draw[thick,dashed] (1.35,2.4) -- (6.35,2.4);
\draw[thick,dashed,gray] (0.8,0)--(2.8,3.5);
\draw[thick,dashed,gray] (1.8,0)--(3.8,3.5);
\draw[thick,dashed,gray] (2.8,0)--(4.8,3.5);
\draw[thick,dashed,gray] (3.8,0)--(5.8,3.5);
\node at (1.7,3.4) (jugh) {$z$};
\node at (0.9,2.5) (jugh) {$\lambdaqft$};
\node at (0.2,1.3) (jugh) {$\frac{\lambdaqft}{2}$};
\draw[thick,->] (0,0) -- (5,0);
\node at (4.7,-0.2) (jugh) {$w$};
\draw[thick,->] (0,0) -- (2,3.5);
\draw[thick,->,blue] (1.8,0) -- (1.8,1);
\draw[thick,->,red] (1.5,1.2) -- (1.5,1.7);
\draw[thick,->,green!50!black] (3.5,1.2) -- (3.5,1.7);
\draw[thick,->,yellow!50!black] (4.5,1.2) -- (4.5,1.7);
\draw[thick,->,blue] (2.5,1.2) -- (2.5,1.7);
\draw[thick,->,yellow!50!black] (5.17,2.4) -- (5.17,3.4);
\node at (0.8,-0.2) (fhnei){$w_{s1}$};
\node at (1.8,-0.2) (fhnei){$w_{s2}$};
\node at (2.8,-0.2) (fhnei){$w_{s3}$};
\node at (3.8,-0.2) (fhnei){$w_{s4}$};
\node at (4.8,0.6) (fhnei){$\ket{\psi_2}$};
\node at (5.6,1.8) (fhnei){$\ket{\phi_2}$};
\node at (6.2,3) (fhnei){$\ket{\psi_4}$};

\tikzset{shift={(-8,-5)}}
\fill[gray!5!white] (0,0)--(2,3.5)--(7,3.5)--(5,0)--(0,0);
\draw[thick,dashed] (0.65,1.2) -- (5.65,1.2);
\draw[thick,dashed] (1.35,2.4) -- (6.35,2.4);
\draw[thick,dashed,gray] (0.8,0)--(2.8,3.5);
\draw[thick,dashed,gray] (1.8,0)--(3.8,3.5);
\draw[thick,dashed,gray] (2.8,0)--(4.8,3.5);
\draw[thick,dashed,gray] (3.8,0)--(5.8,3.5);
\node at (1.7,3.4) (jugh) {$z$};
\node at (0.9,2.5) (jugh) {$\lambdaqft$};
\node at (0.2,1.3) (jugh) {$\frac{\lambdaqft}{2}$};
\draw[thick,->] (0,0) -- (5,0);
\node at (4.7,-0.2) (jugh) {$w$};
\draw[thick,->] (0,0) -- (2,3.5);
\draw[thick,->,green!50!black] (2.8,0) -- (2.8,1);
\draw[thick,->,red] (1.5,1.2) -- (1.5,1.7);
\draw[thick,->,green!50!black] (3.5,1.2) -- (3.5,1.7);
\draw[thick,->,yellow!50!black] (4.5,1.2) -- (4.5,1.7);
\draw[thick,->,blue] (2.5,1.2) -- (2.5,1.7);
\draw[thick,->,red] (2.17,2.4) -- (2.17,3.4);
\node at (0.8,-0.2) (fhnei){$w_{s1}$};
\node at (1.8,-0.2) (fhnei){$w_{s2}$};
\node at (2.8,-0.2) (fhnei){$w_{s3}$};
\node at (3.8,-0.2) (fhnei){$w_{s4}$};
\node at (4.8,0.6) (fhnei){$\ket{\psi_3}$};
\node at (5.6,1.8) (fhnei){$\ket{\phi_3}$};
\node at (6.2,3) (fhnei){$\ket{\psi_1}$};

\tikzset{shift={(8,0)}}
\fill[gray!5!white] (0,0)--(2,3.5)--(7,3.5)--(5,0)--(0,0);
\draw[thick,dashed] (0.65,1.2) -- (5.65,1.2);
\draw[thick,dashed] (1.35,2.4) -- (6.35,2.4);
\draw[thick,dashed,gray] (0.8,0)--(2.8,3.5);
\draw[thick,dashed,gray] (1.8,0)--(3.8,3.5);
\draw[thick,dashed,gray] (2.8,0)--(4.8,3.5);
\draw[thick,dashed,gray] (3.8,0)--(5.8,3.5);
\node at (1.7,3.4) (jugh) {$z$};
\node at (0.9,2.5) (jugh) {$\lambdaqft$};
\node at (0.2,1.3) (jugh) {$\frac{\lambdaqft}{2}$};
\draw[thick,->] (0,0) -- (5,0);
\node at (4.7,-0.2) (jugh) {$w$};
\draw[thick,->] (0,0) -- (2,3.5);
\draw[thick,->,yellow!50!black] (3.8,0) -- (3.8,1);
\draw[thick,->,red] (1.5,1.2) -- (1.5,1.7);
\draw[thick,->,green!50!black] (3.5,1.2) -- (3.5,1.7);
\draw[thick,->,yellow!50!black] (4.5,1.2) -- (4.5,1.7);
\draw[thick,->,blue] (2.5,1.2) -- (2.5,1.7);
\draw[thick,->,blue] (3.17,2.4) -- (3.17,3.4);
\node at (0.8,-0.2) (fhnei){$w_{s1}$};
\node at (1.8,-0.2) (fhnei){$w_{s2}$};
\node at (2.8,-0.2) (fhnei){$w_{s3}$};
\node at (3.8,-0.2) (fhnei){$w_{s4}$};
\node at (4.8,0.6) (fhnei){$\ket{\psi_4}$};
\node at (5.6,1.8) (fhnei){$\ket{\phi_4}$};
\node at (6.2,3) (fhnei){$\ket{\psi_2}$};

\end{tikzpicture}
\caption{Bases for the qu-quarts BB84 protocol.}
\label{fig6}
\end{figure*}

\subsection{Derivation of the Hamiltonian for the qu-quarts scheme}
To have a precise physical description of the qu-quarts scheme, we derive the Hamiltonian (see, Eq.~(\ref{Schroedinger})) for this configuration, which takes into account all possible interactions. To begin with, we analyze the propagation of the classical electromagnetic field in the nonlinear optical fiber by means of the Nonlinear Schr\"{o}dinger Equation~\cite{agrawal2007nonlinear}

\begin{equation}
i\frac{\partial A}{\partial z}=\hat{\beta}A-\gamma\left| A \right|^2 A,
\label{NLSE}
\end{equation}
where $A$ is the complex envelope of the electrical field normalized such that $\left| A \right|^2$ is the optical power in Watts, $\hat{\beta}$ is a linear operator such that $\hat{\beta}e^{-iwt}=-\beta(w)e^{-iwt}$, with $\beta(w)$ the dispersion profile of the fiber, and $\gamma$ the fiber nonlinear coefficient. We propose the solution
\begin{equation*}
A = \sum_{m=1}^{3}\mathbf{A}_{pm}+\sum_{m=1}^{4}\mathbf{A}_{sm},
\end{equation*}
where
\small{
\begin{equation*}
\mathbf{A}_{pm}=\sqrt{P_m}e^{ik_{pm}z-iw_{pm}t}, \quad k_{pm}=\beta(w_{pm}) - \gamma P_m +\sum_{n=1}^{3}2 \gamma P_n,
\end{equation*}}
\begin{equation*}
\mathbf{A}_{sm}=A_{sm}(z)e^{ik_{sm}z-iw_{sm}t}, \quad k_{sm}=\beta(w_{sm})+ \sum_{n=1}^{3}2 \gamma P_n.
\end{equation*}
$\mathbf{A}_{pm}$ correspond to the pumps (under the usual undepleted-pump approximation) and $\mathbf{A}_{sm}$ are small signals located in the qu-quart region. By replacing into Eq.~(\ref{NLSE}) and retaining only the phase-matched QFT processes, we obtain 
\begin{equation}
\label{classical}
i\frac{\partial A_{sm}}{\partial z}=-\kappa \left(A_{s(m-1)}+A_{s(m+1)}\right), 
\end{equation}
for $m=1,2,3,4$, with $A_{s5}=A_{s1}$ and $A_{s0}=A_{s4}$, where $P_1 = P_3 $.

We quantize the variables $A_{sm}$ by proposing the correspondence with quantum operators $\hat{A}_{sm}  = \sqrt{\hbar w_0/T}\hat{a}_{sm}$, where $\hat{a}_{sm}$ is an annihilation operator, $T$ an arbitrary time of measurement and $w_0 = (w_{s1}+w_{s2}+w_{s3}+w_{s4})/4$. Creation operators satisfy the commutation relation $\left[\hat{a}_{sm},\hat{a}^{\dagger}_{sn}
\right] = \delta_{nm}$, and they conform the number operators $\hat{N}_{sm} = \hat{a}^{\dagger}_{sm}\hat{a}_{sm}$, which might be interpreted as the photon quantity at frequency $w_{sm}$ during the time interval $T$. By defining the intensity operator $\hat{I}_{sm}=\hat{A}^{\dagger}_{sm}\hat{A}_{sm}$, we recover the classical relation

\begin{equation*}
\braket{\hat{I}_{sm}}=\frac{\hbar w_0}{T}\braket{\hat{N}_{sm}},
\end{equation*}
where we assume that the qu-quart region bandwidth ($w_{s4}-w_{s1}$) is negligible with respect to $w_0$. If this were not the case, a straightforward correction can be made by considering a frequency-dependent $\gamma$ (i.e., self-steepening); for the sake of clarity, we keep the simpler approximate form. 
We look for a Hamiltonian so that the correspondence principle is satisfied, i.e., the mean values of $\hat{A}_{sm}$ evolve according to the classical Eq.~(\ref{classical})
\begin{equation*}
i\frac{\partial \braket{\hat{A}_{sm}}}{\partial z}=-\kappa \left(\braket{\hat{A}_{s(m-1)}}+\braket{\hat{A}_{s(m+1)}} \right),
\end{equation*}
that is tantamount to say
\begin{equation*}
\left[\hat{A}_{sm},\hat{H}\right]=\kappa \left( \hat{A}_{s(m-1)}+\hat{A}_{s(m+1)}\right).
\end{equation*}
 
Then, we propose the following Hamiltonian
\begin{equation*}
\hat{H} = \kappa \frac{T}{\hbar w_0}\sum_{\mu=1}^4\left(\hat{A}^{\dagger}_{s\mu}\hat{A}_{s(\mu-1)}+\hat{A}^{\dagger}_{s(\mu-1)}\hat{A}_{s\mu} \right),
\end{equation*}
which can be easily verified by keeping in mind the commutation relations $[\hat{A}_{sm},\hat{A}^{\dagger}_{sn}]=\hbar w_0/T\delta_{nm}$. Finally, $\hat{H}$ can be expressed in terms of creation and annihilation operators as
\begin{equation}
\label{Evolution}
\hat{H} = \kappa \sum_{\mu=1}^4\left(\hat{a}^{\dagger}_{s\mu}\hat{a}_{s(\mu-1)}+\hat{a}^{\dagger}_{s(\mu-1)}\hat{a}_{s\mu} \right).
\end{equation}

\subsection{Single-photon propagation}
The quantum state of the complex envelope of the electric field in the qu-quart region is represented by the linear combination of four-dimensional Fock states~\cite{mandel1995optical} $\ket{n_{s1},n_{s2},n_{s3},n_{s4}}$ where $n_{sm}$ is the number of photons at the mode $w_{sm}$. We are interested in having a single-photon at the input-end of the fiber with initial state being either $\ket{w_{s1}}=\ket{1,0,0,0}$, $\ket{w_{s2}}=\ket{0,1,0,0}$, $\ket{w_{s3}}=\ket{0,0,1,0}$ or $\ket{w_{s4}}=\ket{0,0,0,1}$. We may express the evolution of these states after any given propagated distance as

\begin{equation*}
\ket{\psi}(z)=\mu(z) \ket{w_{s1}}+\nu(z) \ket{w_{s2}}+\varrho(z) \ket{w_{s3}}+\varphi(z) \ket{w_{s4}},
\end{equation*}
and obtain precise solutions using the circular relations $\ket{w_{s5}}=\ket{w_{s1}}$ and  $\ket{w_{s0}}=\ket{w_{s4}}$ together with the fact that $\hat{H}\ket{w_{sm}}=\kappa\left(\ket{w_{s(m+1)}}+\ket{w_{s(m-1)}}\right)$.  Then the quantum state of the electric field ($\ket{\psi}$) can be interpreted as the quantum state of the single-photon, where $|\mu|^2$, $|\nu|^2$, $|\varrho|^2$ and $|\varphi|^2$ are the probabilities of measuring its corresponding frequency. 

Using Eq.~(\ref{Schroedinger}) we obtain
\begin{equation*}
\frac{d}{d z}\left(\begin{array}{c} \mu \\ \nu \\ \varrho \\ \varphi \end{array} \right) = i\kappa\left(\begin{array}{cccc} 0 & 1 & 0 & 1\\ 1 & 0 & 1 & 0\\ 0 & 1 & 0 & 1\\ 1 & 0 & 1 & 0 \end{array} \right)\left(\begin{array}{c} \mu \\ \nu \\ \varrho \\ \varphi \end{array} \right),
\end{equation*}
whose solution can be expressed as
\begin{widetext}
\begin{equation}
\label{ququarts}
\left(\begin{array}{c} \mu(z) \\ \nu(z) \\ \varrho(z)\\ \varphi(z) \end{array} \right) = \frac{1}{2}\left( \begin{array}{cccc} \cos(2\kappa z)+1 & i\sin(2\kappa z) & \cos(2\kappa z)-1 & i\sin(2\kappa z)\\ i\sin(2\kappa z) & \cos(2\kappa z)+1 & i\sin(2\kappa z) & \cos(2\kappa z)-1 \\ \cos(2\kappa z)-1 & i\sin(2\kappa z) & \cos(2\kappa z)+1 & i\sin(2\kappa z) \\ i\sin(2\kappa z)& \cos(2\kappa z)-1 & i\sin(2\kappa z)& \cos(2\kappa z)+1 \end{array}\right) \left(\begin{array}{c} \mu(0) \\ \nu(0) \\ \varrho(0)\\ \varphi(0) \end{array} \right).
\end{equation}
\end{widetext}

\subsection{BB84 with qu-quarts}
The scheme is similar to that of the qubit scenario (Fig.~\ref{fig4}) but with four SPD (at frequencies $w_{s1}$, $w_{s2}$, $w_{s3}$ and $w_{s4}$) instead of two. The bases for the qu-quarts scheme are obtained in the same way as before. The $\psi$ basis is used when Alice sends a single-photon directly as it is emitted from the SPS. However, to use the $\phi$ basis, Alice propagates the photon through a $\lambdaqft/2$-long nonlinear fiber before sending it. In Fig.~\ref{fig6} we show the bases for the qu-quarts QKD. They can be easily derived from Eq.~(\ref{ququarts}) and are given by
\begin{equation*}
\ket{\phi_1} = \frac{1}{2}\left(\ket{\psi_1}+i\ket{\psi_2}-\ket{\psi_3}+i\ket{\psi_4} \right),
\end{equation*}
\begin{equation*}
\ket{\phi_2} = \frac{1}{2}\left(i\ket{\psi_1}+\ket{\psi_2}+i\ket{\psi_3}-\ket{\psi_4} \right),
\end{equation*}
\begin{equation*}
\ket{\phi_3} = \frac{1}{2}\left(-\ket{\psi_1}+i\ket{\psi_2}+\ket{\psi_3}+i\ket{\psi_4} \right),
\end{equation*}
\begin{equation*}
\ket{\phi_4} = \frac{1}{2}\left(i\ket{\psi_1}-\ket{\psi_2}+i\ket{\psi_3}+\ket{\psi_4} \right).
\end{equation*}

As in the qubits scheme, Bob choses the basis in which he reads the qu-quarts by deciding whether to propagate the photon ($\phi$) or not ($\psi$) through another $\lambdaqft/2$-long fiber before measuring its frequency. If Alice and Bob choose the same $\phi$ basis, Bob will measure a photon at a different frequency than that of the photon prepared by Alice (Fig.~ \ref{fig6}). This poses no problem in the practical implementation of the protocol, as the frequency measured by Bob is uniquely determined by the frequency sent by Alice.

Security advantages of a qu-quarts BB84 protocol have already been studied~\cite{bechmann2000quantum}. As a very simple proof of the security enhancement, an analysis of the intercept-resend attack can be performed. This simple eavesdropping technique consists in reading the frequency of the photon sent by Alice and resending Bob another photon at the measured frequency. If Alice and Bob chose the $\psi$ basis, this intercept does not produce an error in the communication. However, if they use the $\phi$ basis, then Bob receives a photon of a maximally uncertain frequency. As such, the result of Bob's measurement is completely uncertain and this introduces errors in the communication that can be used to detect the presence of the eavesdropper. For qubits this uncertain measurement produces, in average, $50\%$ of detectable errors; meanwhile for qu-quarts it accounts for $75\%$ of errors. As such, for qubits the Bit Error (BE) introduced by the eavesdropper is $1/4$, meanwhile for qu-quarts the Quart Error (QE) is $3/8$. This simple calculation shows that detection of eavesdropping is easier in the qu-quarts scheme.

\section{Conclusions}
We proposed an original implementation of the BB84 quantum key distribution protocol, based on the frequency uncertainty of single photons in an quantum frequency-translation process in optical fibers. The scheme is expected to be more robust than current BB84 implementations as the frequency of the photon, as opposed to its polarization/phase, is not affected by mechanical and/or thermal fluctuations of the transmitting medium. Detailed calculations to cope with various design considerations were provided and an original extension of the scheme to qu-quarts was presented, expanding the available character space and providing enhanced security.

\bibliographystyle{unsrt}
\bibliography{biblio}

\end{document}